\documentclass[conference]{IEEEtran}
 
\usepackage{cite}
\usepackage{graphics}
\usepackage{epsfig}
\usepackage[linesnumbered,ruled]{algorithm2e}
\usepackage{color}

\usepackage{url}

\newcommand{\etal}{et al.\xspace}
\newcommand{\todo}[1]{}
\renewcommand{\todo}[1]{{\color{red} TODO: {#1}}}

\begin{document}

\title{Impact of Traditional Sparse Optimizations on a Migratory Thread Architecture}

\author{\IEEEauthorblockN{Thomas B. Rolinger\IEEEauthorrefmark{1}\IEEEauthorrefmark{2} and Christopher D. Krieger\IEEEauthorrefmark{2}} \\ 
\IEEEauthorblockA{\IEEEauthorrefmark{1}
University of Maryland, College Park, MD USA
\IEEEauthorblockA{\IEEEauthorrefmark{2}
Laboratory for Physical Sciences, College Park, MD USA} \\
tbrolin@cs.umd.edu, krieger@lps.umd.edu}}

\maketitle

\begin{abstract}
Achieving high performance for sparse applications is challenging due to irregular access patterns and weak locality. These properties preclude many static optimizations and degrade cache performance on traditional systems. 
To address these challenges, novel systems such as the Emu architecture have been proposed. 
The Emu design uses light-weight migratory threads, narrow memory, and near-memory processing capabilities to address weak locality and reduce the total load on the memory system.
Because the Emu architecture is fundamentally different than cache based hierarchical memory systems, it is crucial to understand the cost-benefit tradeoffs of standard sparse algorithm optimizations on Emu hardware.
In this work, we explore sparse matrix-vector multiplication (SpMV) on the Emu architecture.
We investigate the effects of different sparse optimizations such as dense vector data layouts, work distributions, and matrix reorderings.
Our study finds that initially distributing work evenly across the system is inadequate to maintain load balancing over time due to the migratory nature of Emu threads. 
In severe cases, matrix sparsity patterns produce hot-spots as many migratory threads converge on a single resource.
We demonstrate that known matrix reordering techniques can improve SpMV performance on the Emu architecture by as much as 70\% by encouraging more consistent load balancing. This can be compared with a performance gain of no more than 16\% on a cache-memory based system.
\end{abstract}

\begin{IEEEkeywords}
sparse matrix-vector multiplication, optimizations, Emu architecture, performance study
\end{IEEEkeywords}

\section{Introduction}
\label{sec:intro}
Many applications within scientific computing and big-data analytics rely heavily on efficient implementations of sparse linear algebra algorithms, such as sparse matrix-vector multiplication (SpMV).
Examples include conjugate gradient solvers~\cite{CG}, tensor decomposition~\cite{smith2015splatt}, and graph analytics~\cite{GraphAnalytics}.
Unlike dense linear algebra algorithms, sparse kernels present difficult challenges  to achieving high performance on today's common architectures.
These challenges include irregular access patterns and weak locality, which impede static optimizations and efficient cache utilization.
As a result, there has been an abundance of research regarding the design of data structures and algorithms to take advantage of the capabilities of today's systems, which include deep-memory hierarchy architectures and graphics processing units (GPUs)\cite{GPUSpMVFormat}.

Beyond designing data structures and algorithms for sparse kernels that conform to existing systems, there have been efforts to develop novel architectures that would be better suited for sparse algorithms.
One such effort is the Emu architecture~\cite{IA3EMU}, which is a cache-less system centered around light-weight migratory threads and near-memory processing capabilities.
The premise of this architecture is that the challenges posed by irregular applications can be overcome through the use of fine-grained memory accesses that reduce the memory system load by only transferring light-weight thread contexts. The Emu architecture is described in Section \ref{sec:emuArch}.

To determine the efficacy of such a novel architecture, it is insightful to understand how the impact of existing optimizations for sparse algorithms differs between Emu and cache-memory based systems.
To this end, we explore SpMV on the Emu architecture and investigate the effects of traditional sparse optimizations such as vector data layouts, work distributions, and matrix reorderings.
We focus on SpMV as it is one of the most prevalent sparse kernels, is found across a wide range of applications, and exhibits the algorithmic traits that the Emu architecture targets.
Our implementation leverages the standard Compressed Sparse Row (CSR) data format for storing matrices.

This paper's contributions are as follows:
\begin{itemize}
\item We implement a standard CSR-based SpMV algorithm for the Emu architecture with two different data layout schemes for the vectors and two different work distribution strategies.
\item We conduct a performance evaluation of our implementation and the different sparse optimizations across a set of real-world matrices on the Emu Chick system.
\item We find that initially distributing work evenly across the system is inadequate to maintain load balancing over time due to the migratory nature of Emu threads.
\item We demonstrate that traditional matrix reordering techniques can improve SpMV performance on the Emu architecture by as much as 70\% by encouraging sustained load balancing.
On the other hand, we find that the performance gains of the same reordering techniques on a cache-memory based system is no more than 16\%.

\end{itemize}

The rest of this paper is organized as follows: Section \ref{sec:emuArch} provides a brief overview of the Emu architecture.
The details of our CSR-based SpMV implementation and sparse optimizations are presented in Section \ref{sec:impl}.
Section \ref{sec:perfEval} describes our experimental setup and the results of our performance study.
Related work is presented in Section \ref{sec:relatedWork} and we provide concluding remarks in Section \ref{sec:concl}.

\section{Emu Architecture}
\label{sec:emuArch}
The unique aspects of the Emu architecture are migratory threads and memory-side processing. 
These features are designed to improve the performance of code containing irregular memory access patterns~\cite{IA3EMU}. 
Rather than fetching data from memory and bringing it to the location of the computation, the Emu architecture sends or \textit{migrates} the computation to that data. 
This is accomplished by forcing threads to execute on processors that are co-located with accessed data or by sending computation to co-located memory-side processors. 

\subsection{Emu Nodelet Architecture}
\begin{figure}
\centering
\includegraphics[scale=0.35]{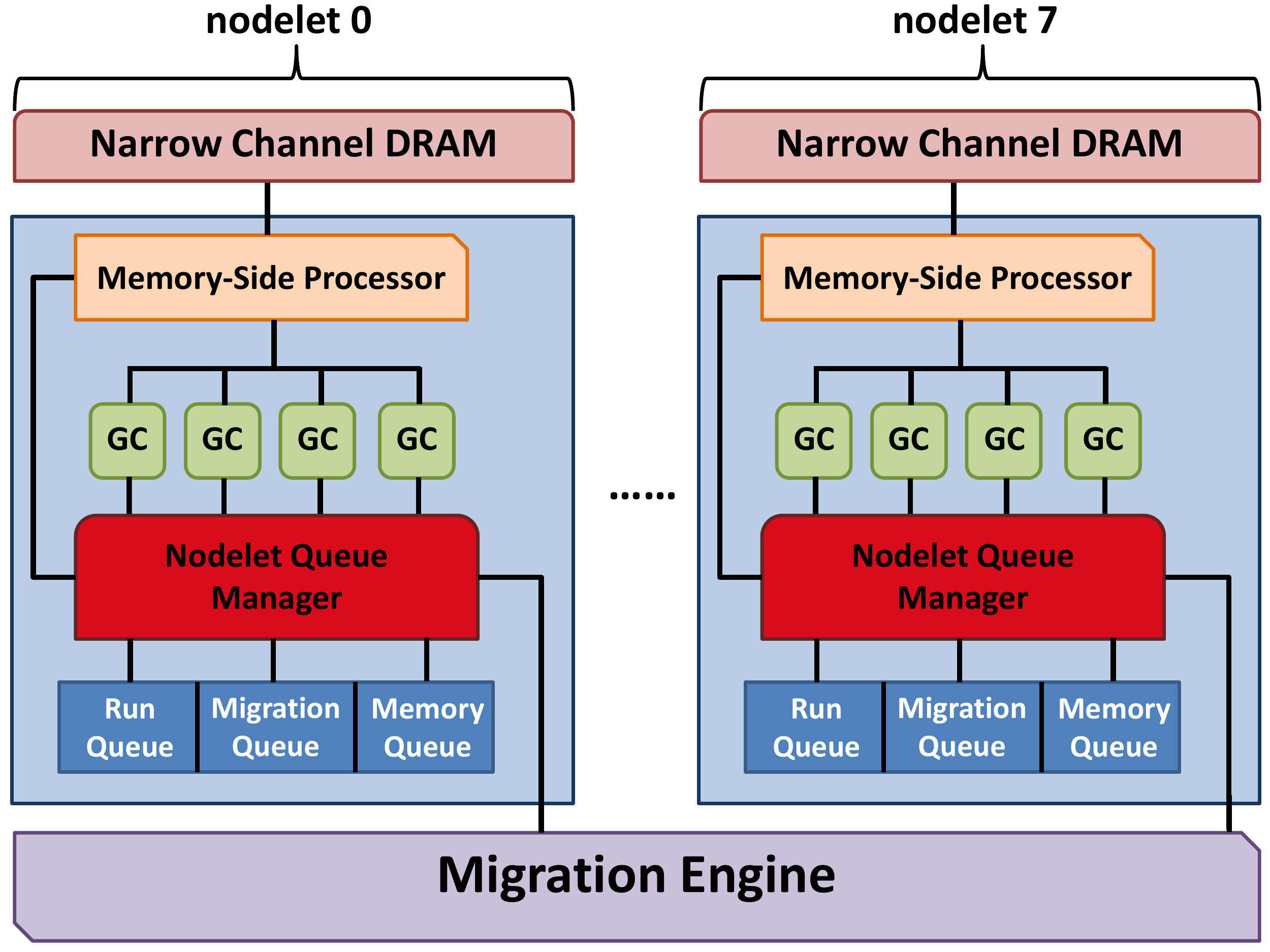}
\caption{A single node in the Emu architecture. ``GC'' refers to a Gossamer Core. There are 8 nodelets within a single node.}
\label{fig:01_EmuArch}
\end{figure}

The basic building block of an Emu system is a \textit{nodelet}, which consists of one or more \textit{Gossamer Cores}, a bank of narrow channel DRAM, and a memory-side processor.
Eight nodelets are combined together to make up a single node.
Figure \ref{fig:01_EmuArch} depicts one such node within the Emu architecture.

A Gossamer Core (GC) is a general purpose, cache-less processing unit developed specifically for the Emu architecture. It supports the execution of up to 64 concurrent light-weight threads.
A GC can issue a single instruction every cycle and each thread on a GC is limited to one active instruction at any given time.
Such restrictions, coupled with the lack of caches, simplifies the logic required by a GC.

%
%

The target applications for the Emu architecture have irregular access patterns and little spatial locality.
Because such applications often require 8 bytes of memory per access, it is inefficient to load 64 byte blocks from main memory. These larger accesses are often required by other architectures due to standard DDR interfaces and cache line sizes.
The narrow channel DRAM within an Emu system is designed to support narrower accesses by using eight 8-bit channels rather than a single, wider 64-bit interface.

When a thread on a GC makes a memory request to a remote address, a \textit{migration} is generated.
A migration involves a GC issuing a request to the Nodelet Queue Manager (NQM) to migrate the thread context to the nodelet that contains the desired data.
The NQM interfaces with the \textit{Migration Engine} (ME), which is the communication fabric that connects multiple nodelets together.
The thread context sits in the migration queue until it is accepted by the ME, at which point it is sent over the ME and is processed by the destination nodelet's NQM.
An executable's compiled code is replicated on each nodelet, so the thread can resume execution without being aware that it was migrated.
By limiting the size of a thread context to roughly 200 bytes~\cite{InitialEmu}, the Emu architecture is able to keep the cost of a migration low.
We have observed that a memory access that requires a migration is roughly 2x slower than a memory access without a migration.

The current Emu architecture supports a range of atomic operations on 64-bit data.
These include add, AND, OR, XOR, min and max.
Atomic operations are handled by the memory-side processor on each nodelet.
For those atomic and store operations that do not return a value to a thread, the memory-side processor can perform the operation on behalf of a thread executing on a different nodelet without generating a migration.
Such operations are referred to as \textit{remote updates}.
A remote update generates a packet that is sent to the nodelet owning the remote memory location. The packet contains the data as well as the operation to be performed.
While remote updates do not return results, they generate acknowledgements that are sent back to the source thread.
A thread cannot migrate until all outstanding acknowledgements have been received. 

Each nodelet has three queues that hold threads in various states.
The run queue consists of threads that are waiting for an available register set on a GC.
The migration queue holds threads, or packets, that are waiting to depart the nodelet.
Packets containing remote updates are held in the memory queue until they can be executed by the memory-side processor.
In the current architecture, the number of active threads on a nodelet can be throttled depending on the availability of resources on that nodelet, including space in these queues.

In this work, we perform our experiments on the Emu Chick system, which consists of 8 nodes connected together by a RapidIO network.
In the current Emu Chick system, each nodelet consists of only one GC clocked at 150MHz and 8GB of narrow channel DDR4 memory clocked at 1600MHz.
The GC and ME on each nodelet are implemented on an Arria 10 FPGA.

\subsection{Data Distribution Support}
There are two basic functions for specifying where allocated memory is placed within the system.
The \texttt{mw\_malloc1dlong} function will allocate an array of 64-bit elements that is distributed cyclically element by element across the available nodelets in the system.
The \texttt{mw\_malloc2d} function allocates an array of pointers that is distributed cyclically element by element across the available nodelets, where each pointer points to a block of co-located memory of a specified size.
While \texttt{mw\_malloc2d} does not directly support variable size blocks, one can achieve such a layout by invoking \texttt{mw\_malloc2d} with block size of 1 and then use standard \texttt{malloc} to allocate the desired block size on each of the nodelets.

\section{Implementation}
\label{sec:impl}
In this section, we provide details regarding our implementation of the Compressed Sparse Row (CSR) storage format and its accompanying SpMV algorithm on the Emu architecture.
We also describe how we achieve different data layouts and work distribution strategies.
For the remainder of this paper, we will refer to a generic sparse matrix \textbf{A} as having $M$ rows, $N$ columns and \textit{NNZ} non-zeros.
We consider $\textbf{Ax} = \textbf{b}$ as the formulation of SpMV, where \textbf{x} and \textbf{b} are dense vectors of length $N$ and $M$, respectively.

\subsection{Compressed Sparse Row Format}
\label{sec:csr}
\begin{figure}
\centering
\includegraphics[trim=0cm 7cm 0cm 0cm, clip, scale=0.35]{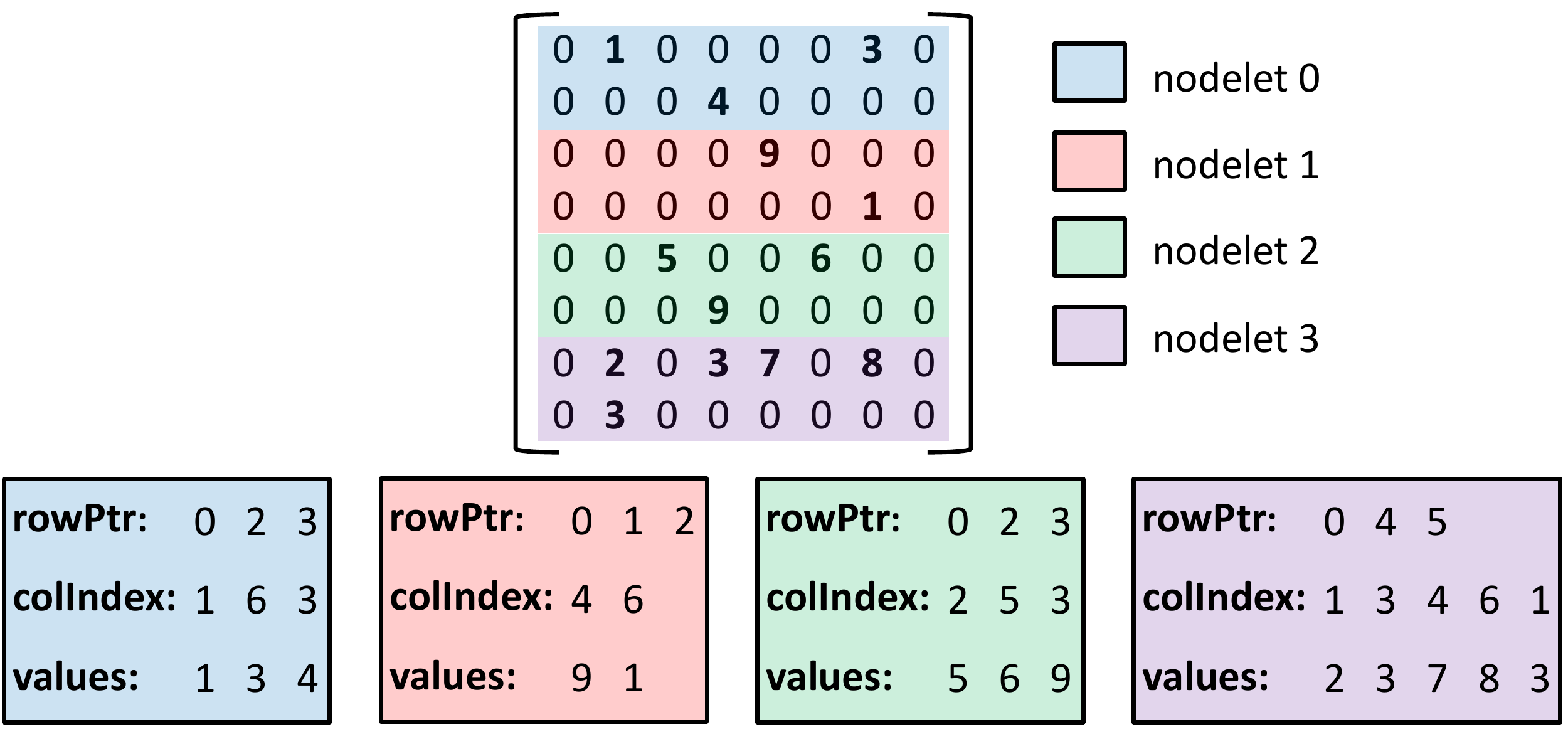}
\caption{Illustration of how a sparse matrix is represented and stored in CSR format across 4 nodelets.}
\label{fig:02_CSRExample}
\end{figure}

We adopt the standard CSR storage format, which stores \textbf{A} as three separate arrays: \textit{values} stores all the non-zero values of \textbf{A}, \textit{colIndex} stores the column indices of all the non-zeros and \textit{rowPtr} stores pointers into \textit{colIndex} that correspond to the start of each row.
We distribute the rows of \textbf{A} across the desired number of nodelets such that each nodelet will have local access to all portions of the CSR arrays it needs to traverse its assigned rows.
An illustration of this distribution for 4 nodelets is shown in Figure \ref{fig:02_CSRExample}, where each nodelet is assigned two rows.
The rows assigned to a given nodelet can then be distributed among the desired number of threads spawned on the nodelet.
Note that each nodelet stores a ``mini'' CSR matrix for its rows with relative row offsets.

%
%

To perform SpMV, we start by spawning a ``parent'' thread on each of the desired nodelets.
Each of these parent threads then spawns the desired number of worker threads to process the rows assigned to the nodelet.
In order to make updates to \textbf{b} during SpMV, each worker thread needs to be aware of the absolute index of its assigned rows, as only relative offsets are stored in the nodelet's \textit{rowPtr} array.
We accomplish this by passing in the absolute row index of the first row assigned to each nodelet when we spawn the parent threads.
Worker threads migrate to and from the nodelets as needed to access elements of \textbf{x} and \textbf{b}.
The rest of the SpMV algorithm is unchanged from the standard CSR implementation.

\subsection{Vector Data Layout}
\label{sec:dataLayout}
While it is possible to enforce only local accesses to the CSR arrays for SpMV, accesses to the vectors are much more challenging to control.
As these memory accesses largely dictate the overall performance of SpMV, it is crucial to address the data layout for \textbf{x} and \textbf{b} on the Emu architecture.

Since the non-zeros in row $i$ of \textbf{A} will only be assigned to a single worker thread, that thread can accumulate the updates to $\textbf{b}[i]$ in a local register and then issue a single store to $\textbf{b}[i]$.
Therefore, fully computing $\textbf{Ax} = \textbf{b}$ only requires $M$ stores to \textbf{b}, which do not require migrations as they are either local writes or remote updates.
On the other hand, there are a total of \textit{NNZ} loads from \textbf{x} required for SpMV, each of which may require a migration.
As it is often the case that \textit{NNZ}~$\gg M$ for most sparse matrices, the layout of \textbf{x} has more bearing on performance than that of \textbf{b}.

We implement two different data layouts for both \textbf{x} and \textbf{b}: cyclic and block.
In a cyclic layout, adjacent elements of a vector are stored on different nodelets in a round-robin fashion such that each consecutive access requires a migration.
For a block layout, contiguous elements of a vector are stored in a fixed-size block on each nodelet.
Assuming a block size of $B$ elements, one migration will be required for every $B$ consecutive accesses.
For our approach, we use the same block size on each nodelet, which is $M/nodelets$ for \textbf{b} and $N/nodelets$ for \textbf{x}, where $nodelets$ is the number of nodelets utilized.

\subsection{Work Distribution Strategies}
\label{sec:workDistr}
We explore two different strategies for distributing work across the nodelets: one that only considers the number of rows in \textbf{A} and one that also factors in the number of non-zeros on each row.
For the row-based approach, we evenly distribute the rows of \textbf{A} to each nodelet and then further divide those blocks of rows among the worker threads utilized by each nodelet.
When using the block layout for \textbf{b}, the block size is equal to the number of rows assigned to each nodelet via the row distribution strategy.
Therefore, the worker threads on a nodelet will have local access to the elements of \textbf{b} that need to be updated.

While each nodelet may receive the same number of rows via the row approach, the sparsity pattern of the matrix may result in some worker threads being assigned a significantly different number of non-zeros than others.
Since the number of non-zeros given to a nodelet largely dictates its work-load for SpMV, this can lead to load imbalance.
To mitigate this, the non-zero approach distributes the rows of \textbf{A} to each worker thread such that the total number of non-zeros assigned to each thread is roughly the same.
We achieve this by iterating over \textit{rowPtr} and accumulating rows until the threshold of $NNZ / threads$ is met, where $threads$ is the total number of worker threads used across all of the nodelets.
For matrices with very irregular sparsity patterns, this can result in a given nodelet being assigned a significantly different number of rows than another.
In such cases, a block layout for \textbf{b} no longer guarantees that the required elements of \textbf{b} will be local to each nodelet.

\section{Performance Evaluation}
\label{sec:perfEval}
In order to understand the impact of traditional sparse optimizations on a migratory thread architecture, 
we evaluated our SpMV implementation on the Emu Chick system across a range of real-world matrices.
In this section, we describe our experimental setup and then present the results of several different experiments that evaluate the sparse optimizations described in Section \ref{sec:impl}.

\subsection{Experimental Setup}
\label{sec:expSetup}
For our experiments, we ran on a single node of the Emu Chick system, as described in Section \ref{sec:emuArch}, and used version 18.04.1 of the Emu toolchain.
Multi-node execution on the Emu Chick hardware was not reliable enough to conduct our tests, so we limit our experiments to a single node and leave multi-node tests for future work.
We utilize all 8 nodelets on a node and leverage 64 worker threads per nodelet.

We executed our SpMV implementation across a suite of 40 different matrices.
In the following sections, we focus our evaluation on the matrices shown in Table \ref{tab:matrices}, which are representative of the suite and highlight the most interesting performance characteristics.
All matrices were obtained from the University of Florida Sparse Matrix Collection~\cite{MatrixMarket} with the exception of rmat, which is an RMAT graph that was generated with RMAT a, b and c parameters of 0.45, 0.22 and 0.22, respectively~\cite{parmat}.
For the symmetric matrices, we store and operate on the entire matrix rather than just the upper or lower triangular matrix.
All results reported are the average of 10 trials.

\begin{table}
\renewcommand{\arraystretch}{1.0}
\caption{Matrices}
\label{tab:matrices}
\centering
\begin{tabular}{|c|c|c|c|c|}
\hline
\textbf{Name} & \textbf{Dimensions} & \textbf{Non-Zeros} & \textbf{Density} & \textbf{Symmetric}\\
\hline
ford1 & 18k x 18k & 100k & 2.9E-04 & Yes \\
cop20k\_A & 120k x 120k & 2.6M & 1.79E-04 & Yes \\
webbase-1M & 1M x 1M & 3.1M & 3.11E-06 & No \\
rmat & 445k x 445k & 7.4M & 3.74E-05 & No\\
nd24k & 72k x 72k & 28.7M & 5.54E-03 & Yes \\
audikw\_1 & 943k x 943k & 77.6M & 8.72E-05 & Yes \\
\hline
\end{tabular}
\end{table}

\subsection{Cyclic Versus Block Vector Data Layout}
\label{sec:cyclicVsBlock}
\begin{figure}
\centering
\includegraphics[trim={0cm 2.5cm 0 0},clip,scale=0.36]{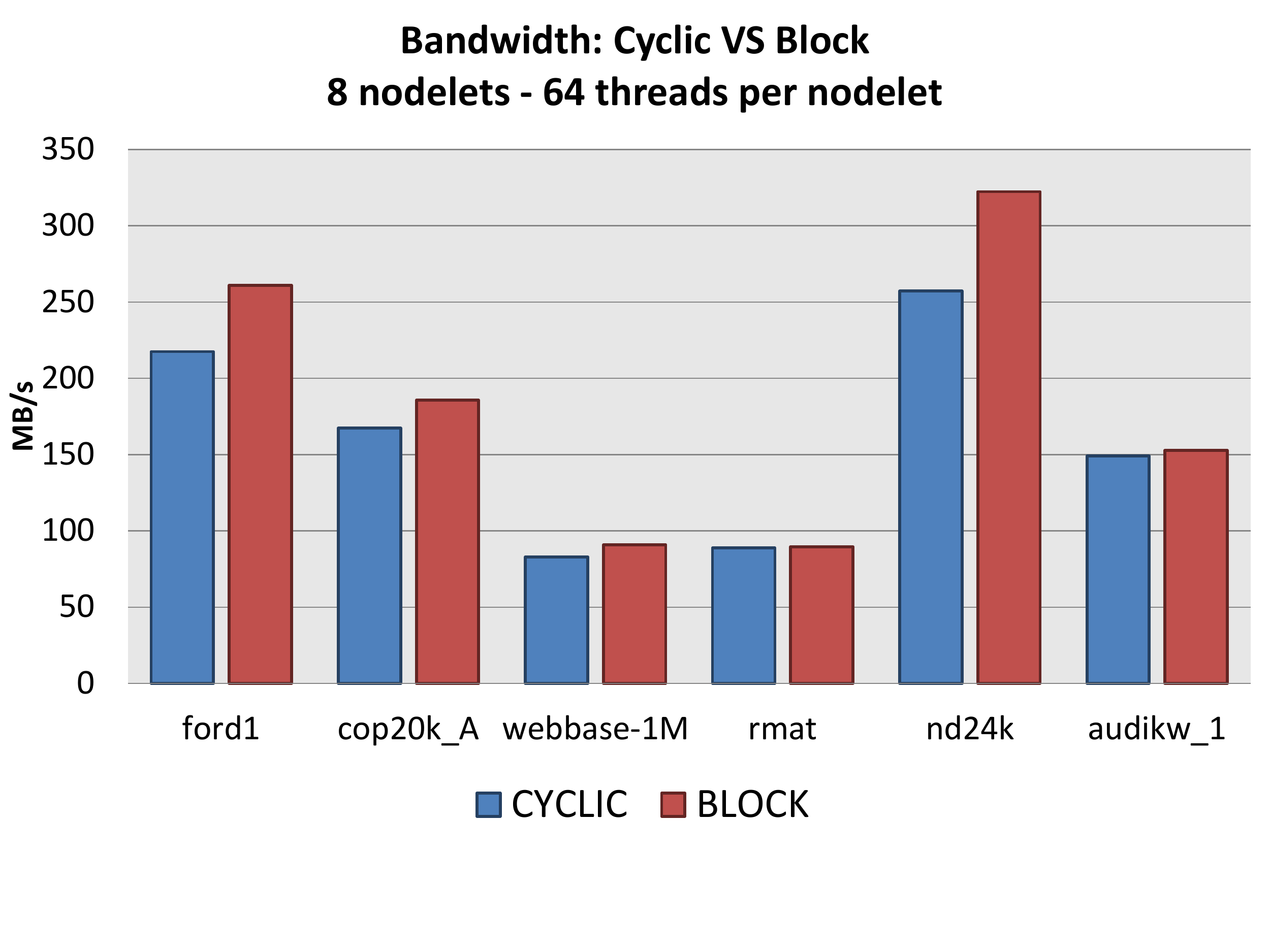}
\caption{Bandwidth (MB/s) for SpMV across the matrices when using a cyclic and block data layout for \textbf{x} and \textbf{b}. Higher bars represent better performance.}
\label{fig:04_blockVsCyclic_BW}
\end{figure}

%
Figure \ref{fig:04_blockVsCyclic_BW} shows the achieved SpMV bandwidth for the cyclic and block vector layouts across the different matrices.
A row-based work distribution strategy was employed for these results.
The block layout outperforms the cyclic layout on each matrix, achieving up to 25\% more bandwidth.

%
%
%
%
\begin{figure}
\centering
\includegraphics[scale=0.30]{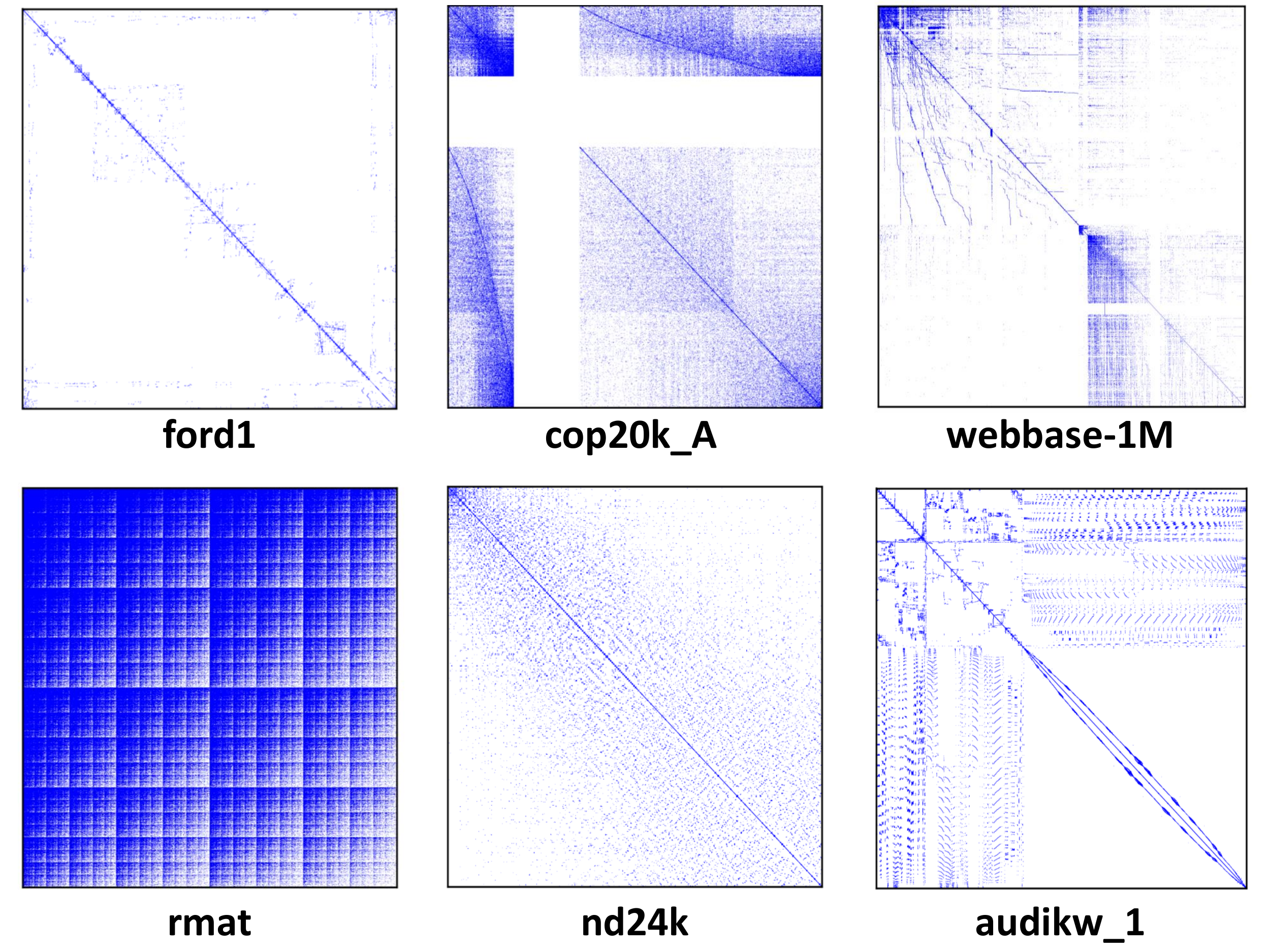}
\caption{Spy plots for the matrices in Table \ref{tab:matrices}. Blue (dark) dots represent non-zeros.}
\label{fig:03_Spyplots}
\end{figure}

\begin{figure}
\centering
\includegraphics[trim={4cm 0 0 0},clip,scale=0.4]{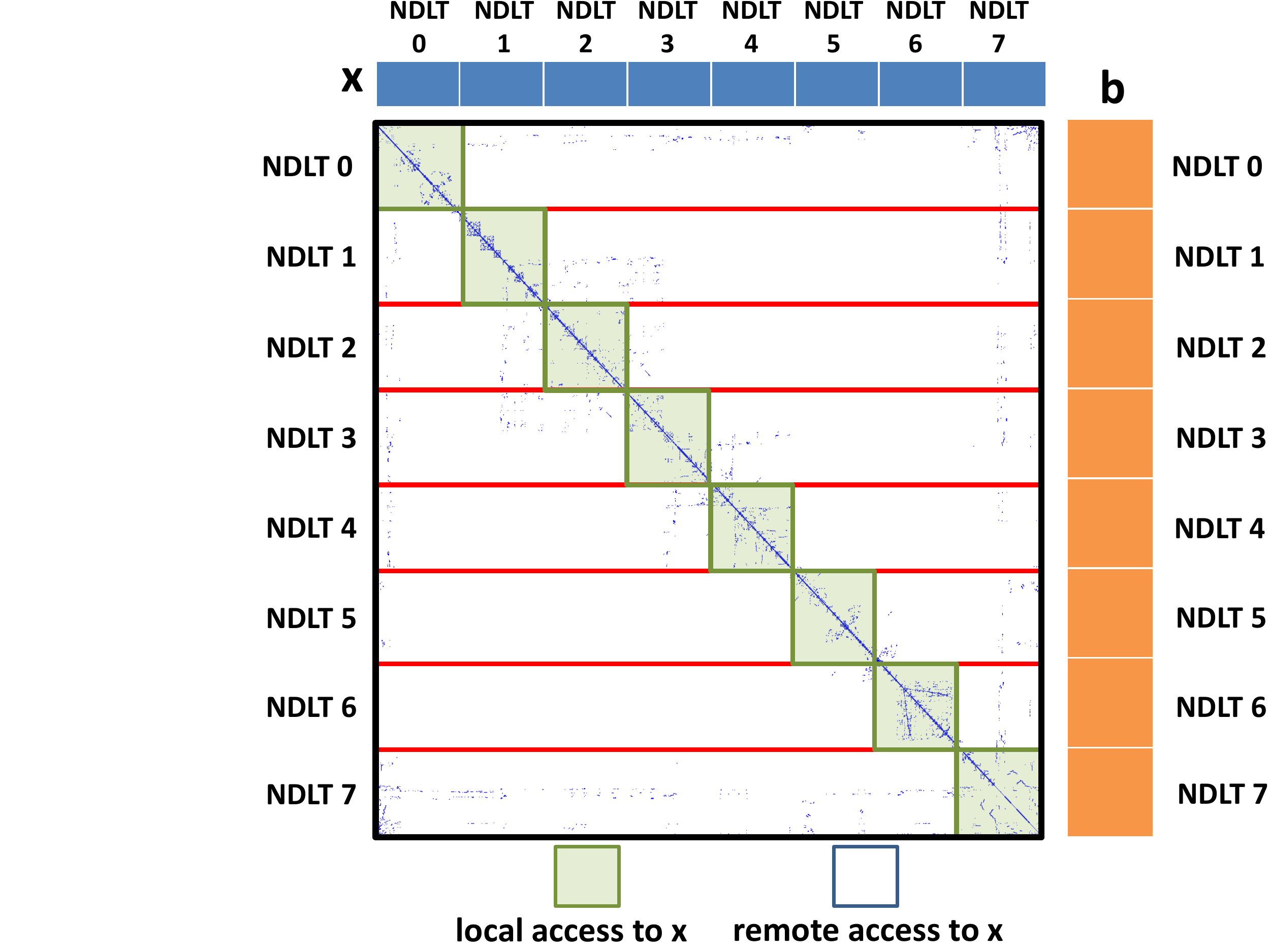}
\caption{An illustration of how the ford1 matrix from Figure \ref{fig:03_Spyplots} and the \textbf{x} and \textbf{b} vectors are distributed across 8 nodelets via the row-based work distribution approach with a block layout for \textbf{x} and \textbf{b}. Red (dark) lines indicate the blocks of rows assigned to each nodelet. Blue (dark) dots within shaded boxes represent non-zeros that require local access to \textbf{x} while those outside of the shaded boxes represent non-zeros that require migrations to access \textbf{x}.}
\label{fig:05_ford1Partitions}
\end{figure}

We can study the sparsity patterns of the matrices, as shown in Figure \ref{fig:03_Spyplots}, to understand why the block layout performs better than the cyclic layout.
A particular sparsity pattern that offers significant benefits for the block layout is one in which a majority of the non-zeros are clustered around the main diagonal of the matrix.
Since the matrices in Table \ref{tab:matrices} are square, the length of both \textbf{x} and \textbf{b} is equal to the number of rows in \textbf{A}.
With the row-based work distribution, the number of rows assigned to each nodelet is equal to the block size used for the vectors.
For a matrix where the non-zeros are clustered on the main diagonal, this means that very few migrations are incurred when accessing \textbf{x}.
Figure \ref{fig:05_ford1Partitions} illustrates this for the ford1 matrix across 8 nodelets.
As can be seen, a majority of the non-zeros are contained in the shaded boxes, which represent local accesses to \textbf{x}.
If \textbf{x} were distributed in a cyclic layout, then we could not exploit such a sparsity pattern, as consecutive accesses to \textbf{x} would cause migrations.
Indeed, we observe that the block layout generates 1.42x -- 6.3x fewer migrations than the cyclic layout across the matrices in Table \ref{tab:matrices}.

For the remainder of the results, we will assume a block data layout for \textbf{x} and \textbf{b}.

%
%

\subsection{Row Versus Non-zero Work Distribution}
\label{sec:rowVSNonzero}
\begin{figure}
\centering
\includegraphics[trim={0.2cm 2.5cm 0 0},clip,scale=0.35]{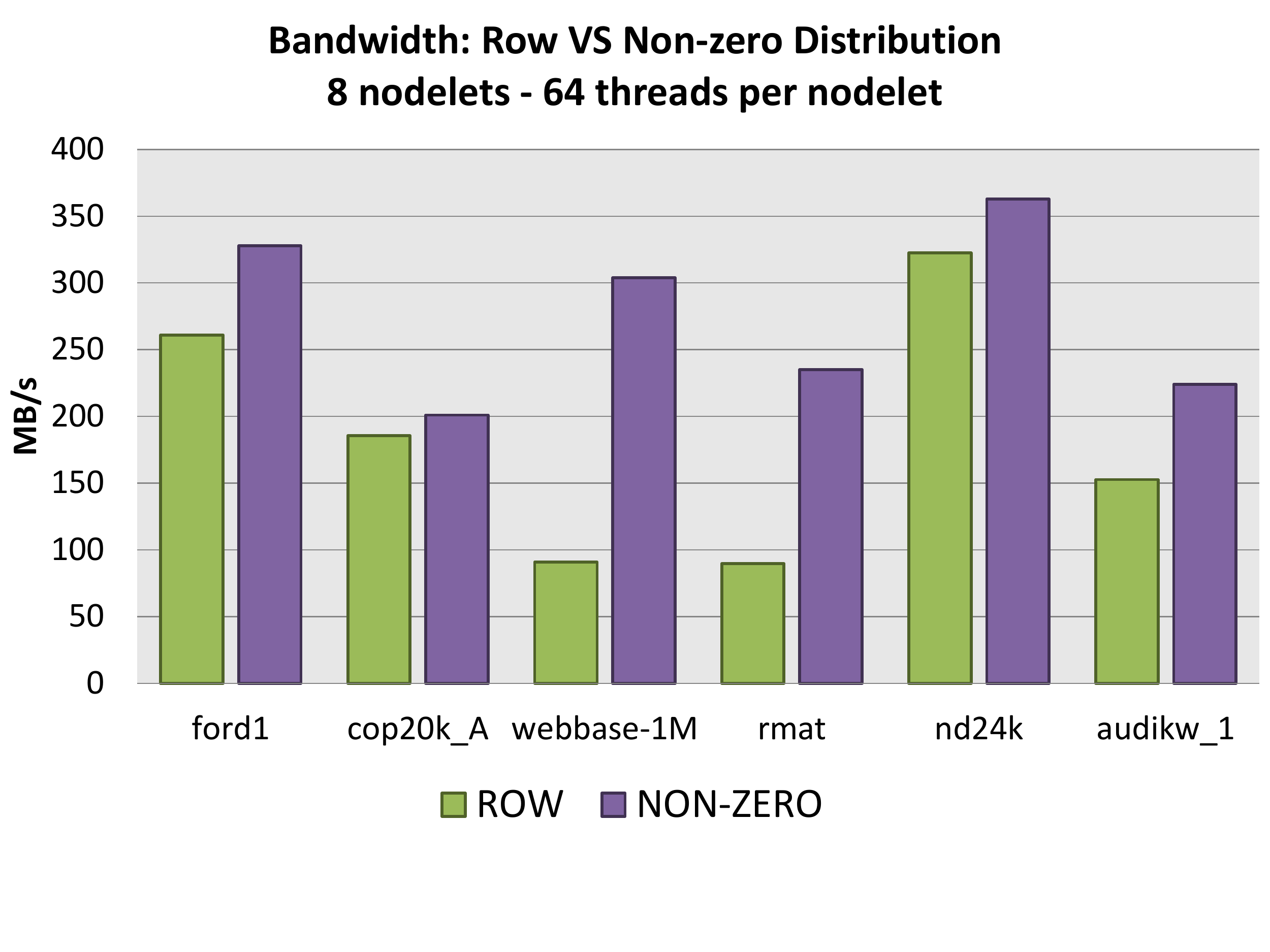}
\caption{Bandwidth (MB/s) for SpMV across the matrices when using a row and non-zero work distribution. Higher bars represent better performance.}
\label{fig:06_RowVSNonzero_BW}
\end{figure}

Figure \ref{fig:06_RowVSNonzero_BW} presents the SpMV bandwidth results of the row and non-zero work distribution strategies across the matrices.
We observe that the non-zero approach consistently outperforms the row approach, achieving as much as 3.34x more bandwidth.
As described in Section \ref{sec:workDistr}, the row-based approach can lead to severe work imbalances by assigning a block of rows to a given nodelet with a significantly different number of non-zeros than other nodelets.

%
%
%
%
\begin{figure}
\centering
\includegraphics[trim={0cm 2.5cm 0 0},clip,scale=0.36]{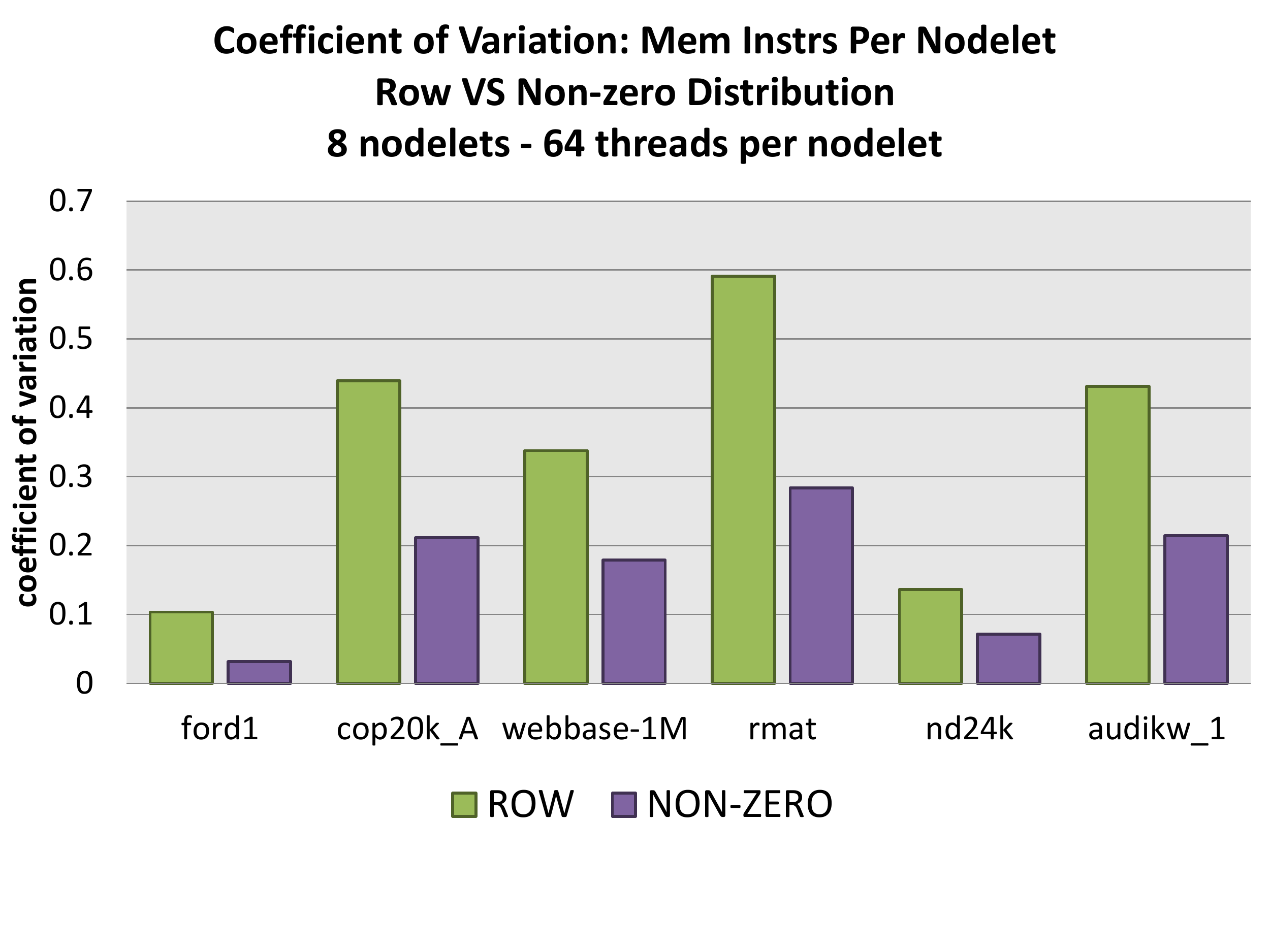}
\caption{Coefficient of variation for the memory instructions executed per nodelet when using a row and non-zero work distribution. Lower bars represent better performance.}
\label{fig:07_RowVSNonzero_MemIPC}
\end{figure}

To quantify the performance advantage of the non-zero work distribution, Figure \ref{fig:07_RowVSNonzero_MemIPC} shows the coefficient of variation (CV) for the number of memory instructions executed by each nodelet.
The CV is the standard deviation divided by the mean and is a measure of relative variability.
A low memory instruction CV indicates that each nodelet completed roughly the same amount of work, as SpMV is considered to be memory-bound. 
Indeed, we observe a significantly lower CV for the non-zero approach across all of the matrices.

While the work is distributed more evenly with the non-zero approach and we observe higher bandwidth, the non-zero strategy incurs an average of 1.69x more migrations than the row-based approach.
This is because the row strategy coupled with a block layout for the vectors works to minimize migrations, especially for matrices with a dense main diagonal.
However, the non-zero distribution does not necessarily assign equally sized blocks of rows to each nodelet.
The result is that the block data layout for the vectors is less successful at minimizing migrations as a nodelet's assigned rows and non-zeros are not necessarily aligned with the block partitions of \textbf{x} and \textbf{b}.

Despite incurring more migrations, the non-zero approach offers better performance than the row-based approach.
This suggests that the penalty of migrations can be offset by more uniform work distribution and load balancing.
We discuss this topic in more detail in the next section.
For the remainder of the results, we assume the use of the non-zero work distribution strategy.

\subsection{Hardware Load Balancing}
\label{sec:hwLoad}
On a traditional cache-memory based system, both memory access locality and hardware load balancing for SpMV can be controlled by distributing the non-zeros among the threads and binding the threads to hardware resources such as cores.
However, the Emu architecture differs because threads cannot be isolated to specific hardware resources, such as a Gossamer Core, due to their migratory behavior.
To bind Emu threads to cores, one would need to only read from local memory, avoiding migrations entirely.
At the other extreme, despite best efforts to initially lay out and distribute work evenly across the nodelets, it is possible for all of the threads to migrate to a single nodelet and oversubscribe that nodelet's resources.
In general, the layout of an application's data structures across the nodelets as well as its memory access pattern are what determine the load balancing of the hardware.

Consider the cop20k\_A matrix as shown in Figure \ref{fig:03_Spyplots}.
Regardless of how the rows are distributed among the nodelets, a large portion of the total non-zeros require access to elements of \textbf{x} that all reside on the same nodelet.
Specifically, 25\% of the total non-zeros in the matrix require access to elements of \textbf{x} that are stored on nodelet 0.
Within the Emu architecture, this results in a majority of the threads all migrating to the same nodelet at roughly the same time.

\begin{figure}
\centering
\includegraphics[scale=0.36]{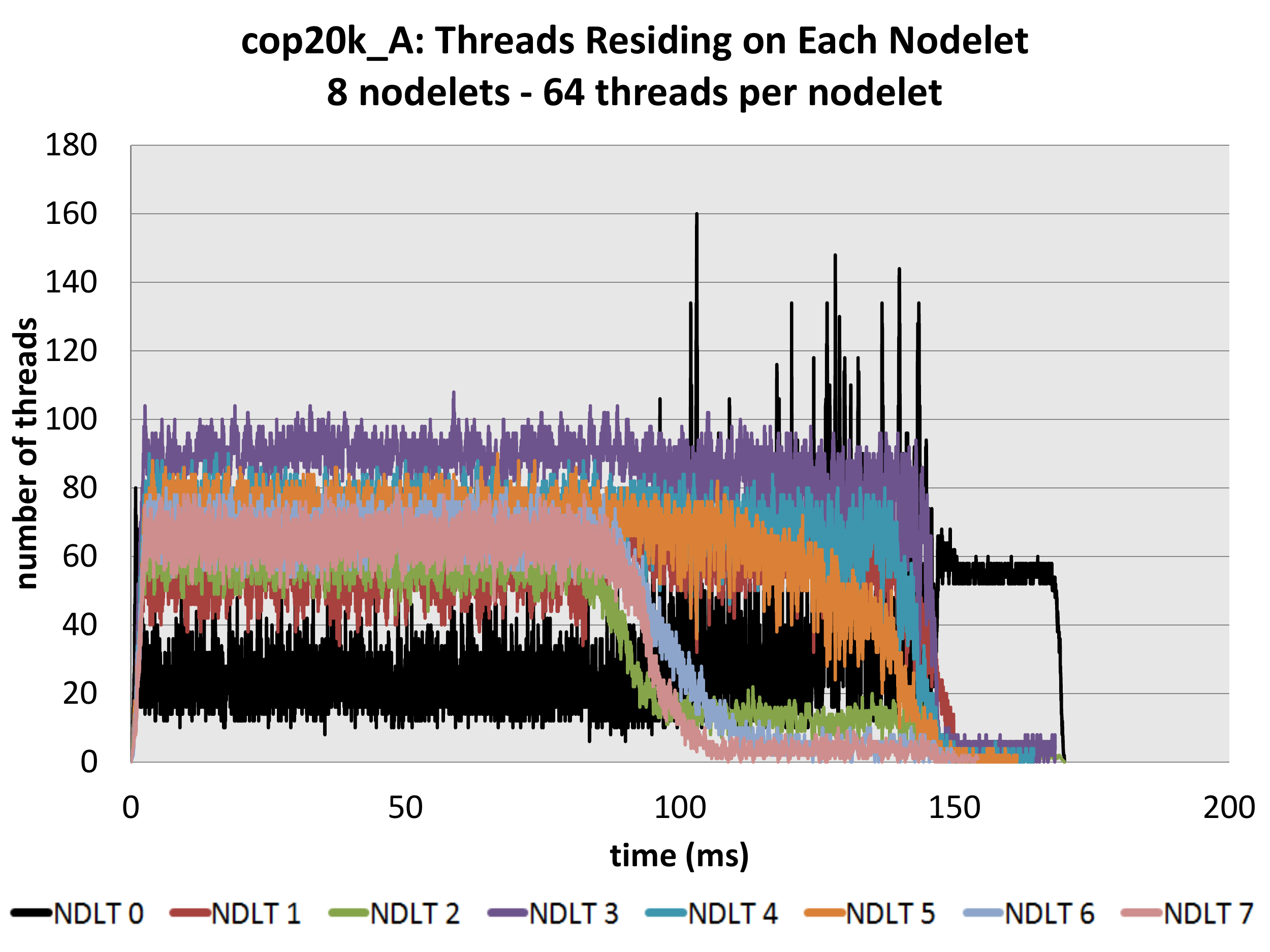}
\caption{The total number of threads on each nodelet during the execution of SpMV on the cop20k\_A matrix. These results include threads executing on the nodelets as well as those waiting in the run queues.}
\label{fig:09_ActiveThreads_Cop_64}
\end{figure}

The particular load imbalance scenario for the cop20k\_A matrix is shown in Figure \ref{fig:09_ActiveThreads_Cop_64}, where we monitor the total number of threads residing on each nodelet during the SpMV execution, including those executing and those waiting in the run queues.
By the time 150ms have elapsed, all of the nodelets except for nodelet 0 have ceased significant activity, indicating that there is a clear load imbalance of the system resources.
However, as described above, we would expect to see higher than average activity on nodelet 0 due to all of the threads requiring access to elements of \textbf{x} stored on nodelet 0.
Instead, the number of threads on nodelet 0 is, on average, 32 while the other nodelets maintained between 53 and 75 threads on average.

%
%
%
%

To understand this behavior, we observed the sizes of the migration queue on each nodelet over the SpMV execution.
We found that nodelet 0 experiences an immediate surge of packets into its migration queue.
This is due to a majority of the worker threads migrating to nodelet 0 to access \textbf{x} and then requiring a migration back to their parent nodelet to access their CSR arrays.
Nodelet 1 also exhibits noticeable activity, as it too holds elements of \textbf{x} that are required by a large number of non-zeros residing on other nodelets.
In the current Emu architecture, thread activity on a nodelet is throttled based on the nodelet's available resources, which includes space in the migration queue.
Because the migration queue on nodelet 0 is immediately filled nearly to capacity, the nodelet reduces the number of threads that can be executed.
It is not until the other nodelets approach completion of their work that the migration queue on nodelet 0 starts to empty out and thread activity increases.
We note that running SpMV on the cop20k\_A matrix with fewer threads per nodelet provides better load balancing by reducing the pressure on the migration queue, and thus, allowing for more threads to be active on nodelet 0.
This suggests that as more nodelets and threads are present as a system is scaled up, load balancing issues due to thread migration hot-spots will require attention.

\subsection{Matrix Reordering}
\label{sec:matReorder}
\begin{figure}
\centering
\includegraphics[trim=0cm 4.5cm 12cm 0cm, clip, scale=0.4]{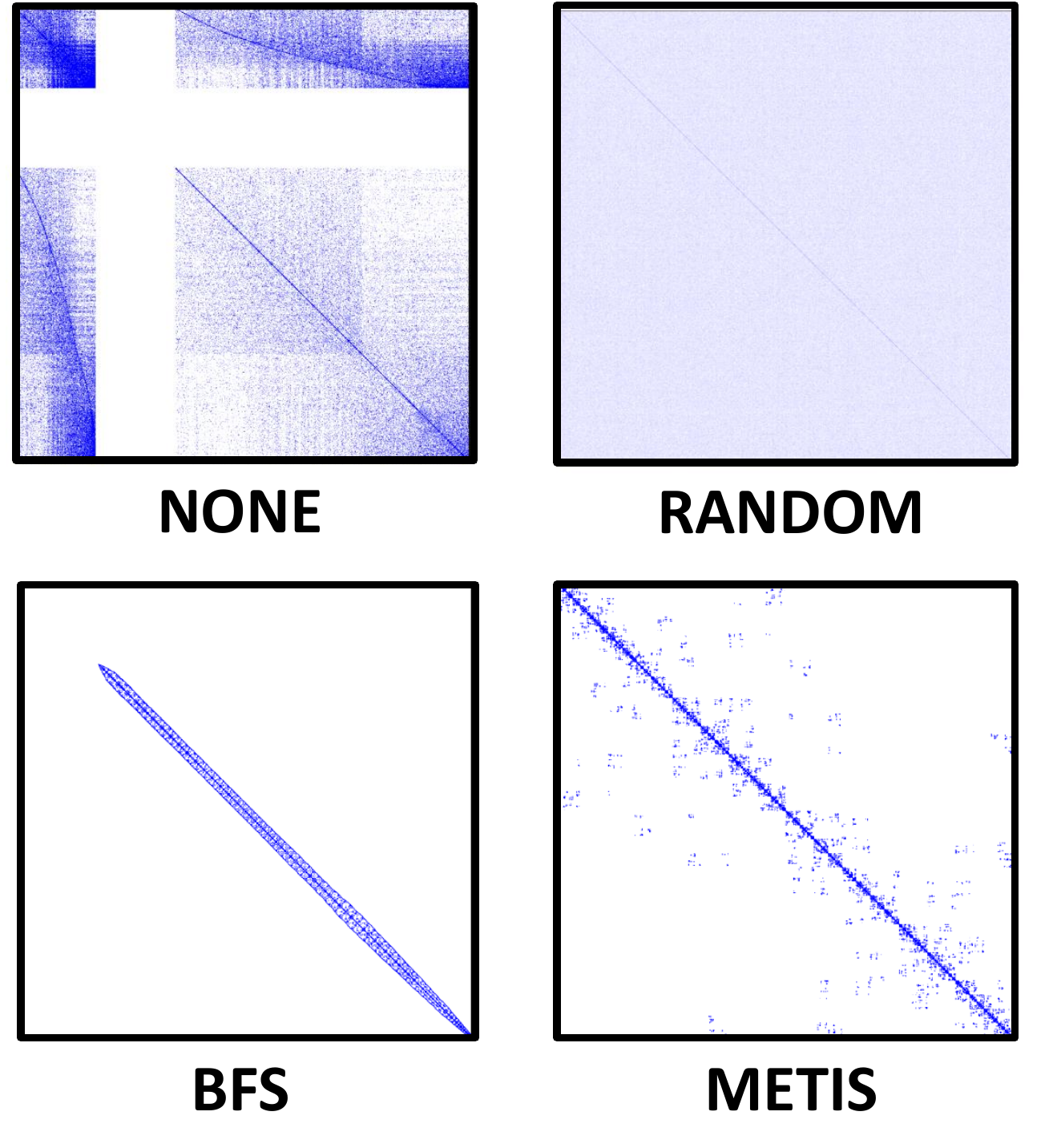}
\caption{The original cop20k\_A matrix, denoted as NONE, and the resulting spy plots when applying the random, BFS and METIS reordering algorithms.}
\label{fig:12_Cop20K_A_Reorderings}
\end{figure}

As shown in the previous section, the sparsity pattern of a given matrix can have profound impacts on SpMV performance, despite efforts to properly lay out data structures and distribute work evenly to all of the threads.
There are many known techniques to reorder the non-zeros of a matrix in order to improve locality and data reuse on traditional cache-memory based systems.
We investigated whether these existing reordering algorithms could offer similar performance benefits on Emu system as well as mitigate potential hardware load balancing issues.
We focus on the following reordering techniques: Breadth First Search (BFS)~\cite{BFS}, METIS~\cite{METIS} and Random.
A random reordering performs a random permutation of the matrix rows and columns using a Fisher-Yates shuffle.
As an example, Figure \ref{fig:12_Cop20K_A_Reorderings} shows the original cop20k\_A matrix and the results of the three reordering algorithms.

%
%
%
%
\begin{figure}
\centering
\includegraphics[trim={0.2cm 2.25cm 0 0},clip,scale=0.35]{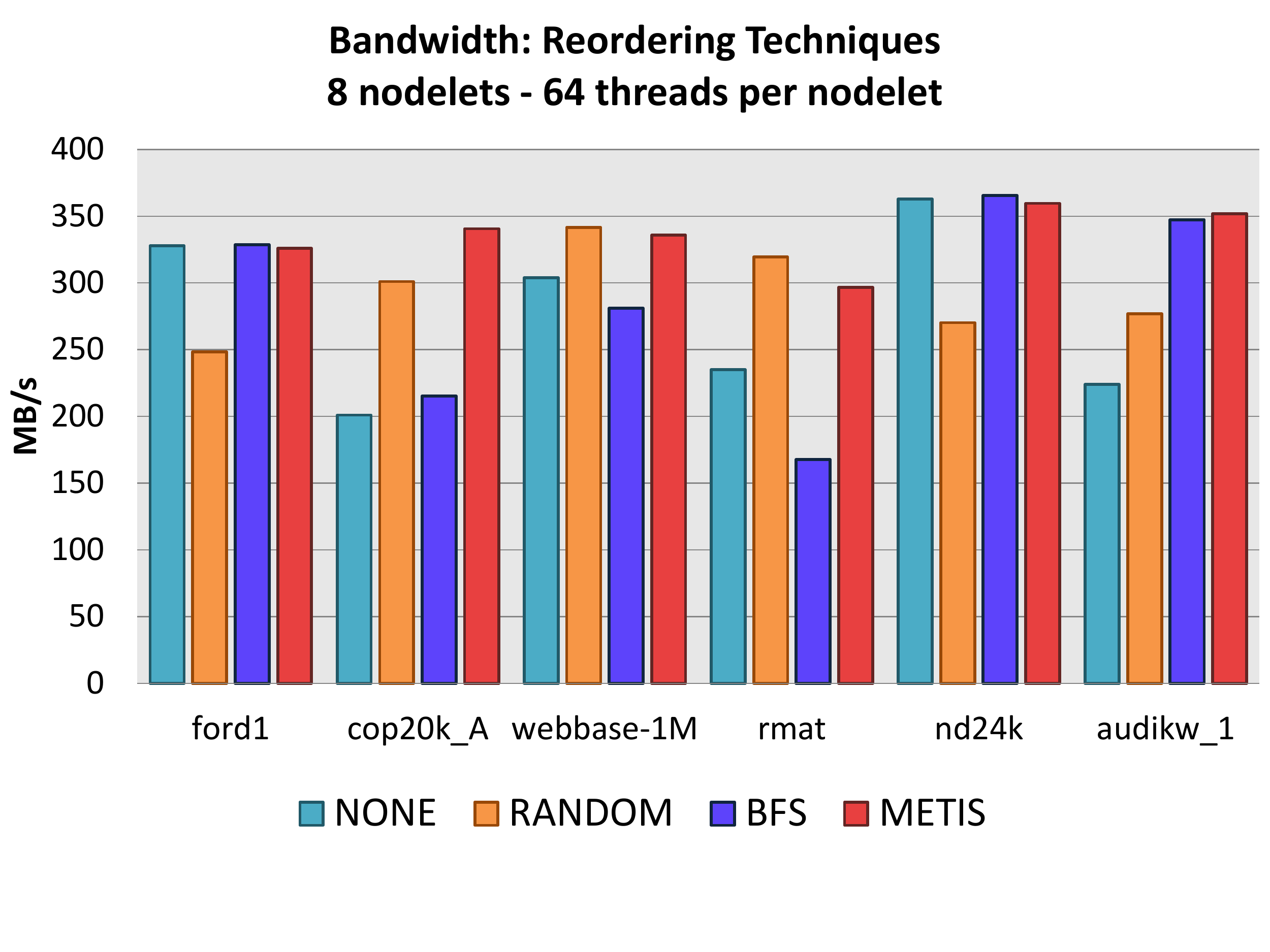}
\caption{Bandwidth (MB/s) of SpMV for the different reordering techniques across the matrices. Higher bars represent better performance. NONE refers to the original matrix.}
\label{fig:13_Reordering_EMUBW}
\end{figure}

Figure \ref{fig:13_Reordering_EMUBW} presents the achieved SpMV bandwidth for the different reordering techniques across the matrices, where NONE refers to the original matrix.
We find that BFS and METIS generally offer the best performance, achieving up to 70\% more bandwidth than the original matrix.
BFS and METIS both attempt to move the non-zeros towards the main diagonal and they tend to put an equal number of non-zeros on each row.
Having the non-zeros clustered on the main diagonal allows us to exploit the block data layout of the vectors and reduce the total amount of migrations, as described in Section \ref{sec:cyclicVsBlock}.
Since BFS and METIS tend to produce balanced rows, both of the work distribution strategies achieve roughly the same outcome: each nodelet is assigned an equal number of rows and the total number of non-zeros assigned to each nodelet is roughly the same.
Therefore, these reordering techniques allow us to maintain an equal amount of activity on each nodelet and mitigate hot-spots by encouraging threads to be ``pinned'' to their parent nodelet.

\begin{figure}
\centering
\includegraphics[scale=0.35]{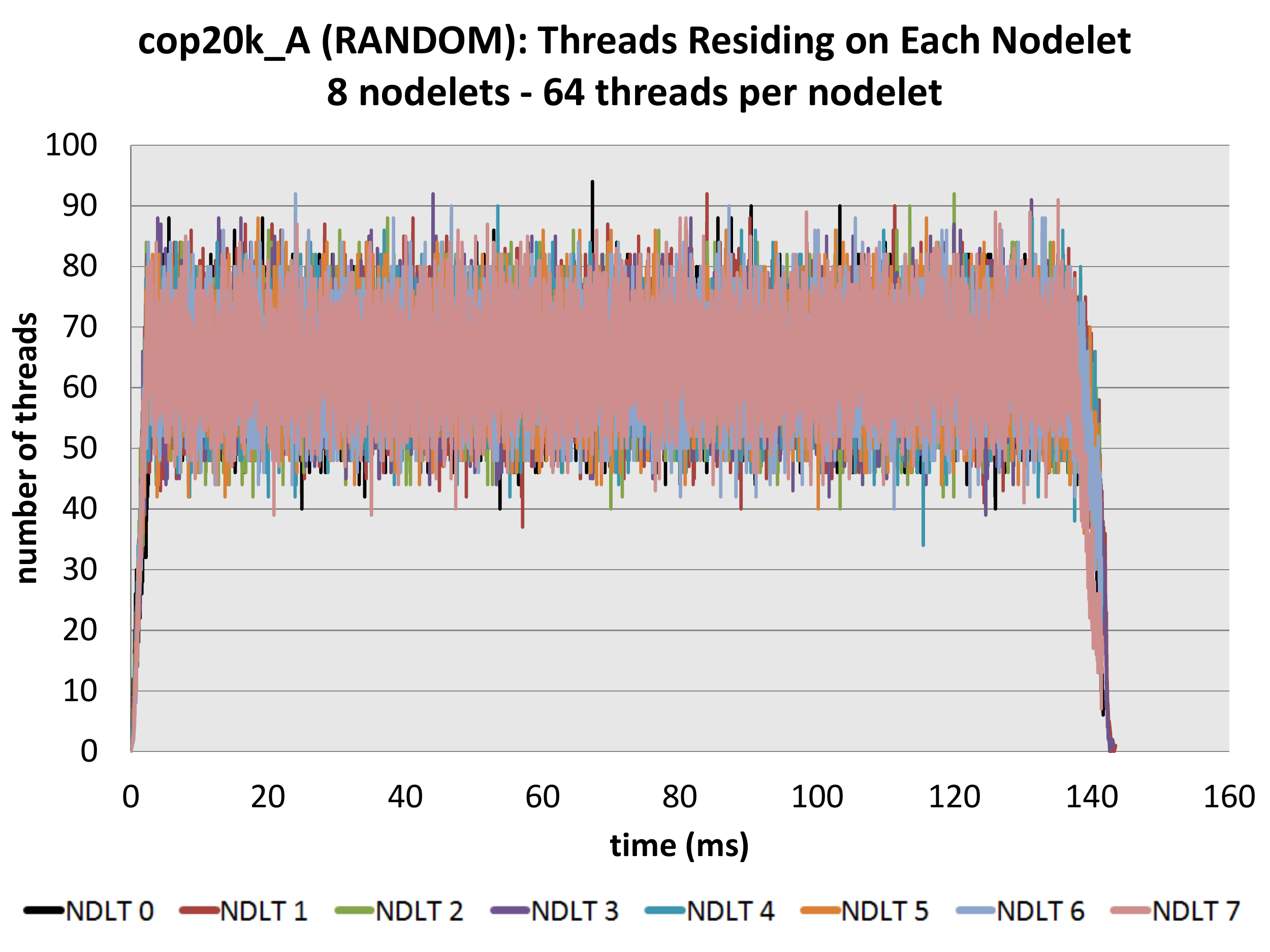}
\caption{The total number of threads on each nodelet during the execution of SpMV on the cop20k\_A matrix when randomly reordered. These results include threads executing on the nodelets as well as those waiting in the run queues.}
\label{fig:15_ActiveThreads_Cop_RANDOM}
\end{figure}

However, an interesting result from Figure \ref{fig:13_Reordering_EMUBW} is that a random reordering can offer up to a 50\% increase in bandwidth over the original matrix, and in some cases, outperform BFS or METIS.
The random reordering has the effect of also producing balanced rows by uniformly spreading out the non-zeros rather than clustering them together.
As one would expect, this results in more migrations than the other techniques, but it provides a natural hot-spot mitigation for SpMV.
This is because it is very unlikely that a majority of the threads would all converge onto the same nodelet at the same time.
We observe that such an effect is similar to the distributed randomized algorithm for packet routing proposed by Valiant~\cite{Valiant}, which prevents multiple packets from being sent across the same wire at the same time.
As alluded to at the end of Section \ref{sec:rowVSNonzero}, the cost of extra migrations can be overcome by better load balancing.
Indeed, we can see the improvement in load balancing achieved by the random reordering in Figure \ref{fig:15_ActiveThreads_Cop_RANDOM}, which tracks the total number of threads on each nodelet during SpMV for the cop20k\_A matrix.

%
%

The results from Figure \ref{fig:13_Reordering_EMUBW} highlight two approaches for achieving hardware load balancing on the Emu architecture: (1) assign an equal amount of work to each nodelet and lay out the data so that threads rarely need to migrate off of their parent nodelet, and (2) assign an equal amount of work to each nodelet and lay out the data so that threads will rarely converge onto the same nodelet at the same time.
The first approach, as achieved by BFS and METIS, attempts to enforce the original intent of the data layout and work distribution, and generally offers the best performance.
The second approach, as achieved by the random reordering, incurs more migrations but can be useful due to the minimal amount of work required to perform the reordering.

%
%
\begin{figure}
\centering
\includegraphics[trim={0cm 2cm 0 0},clip,scale=0.35]{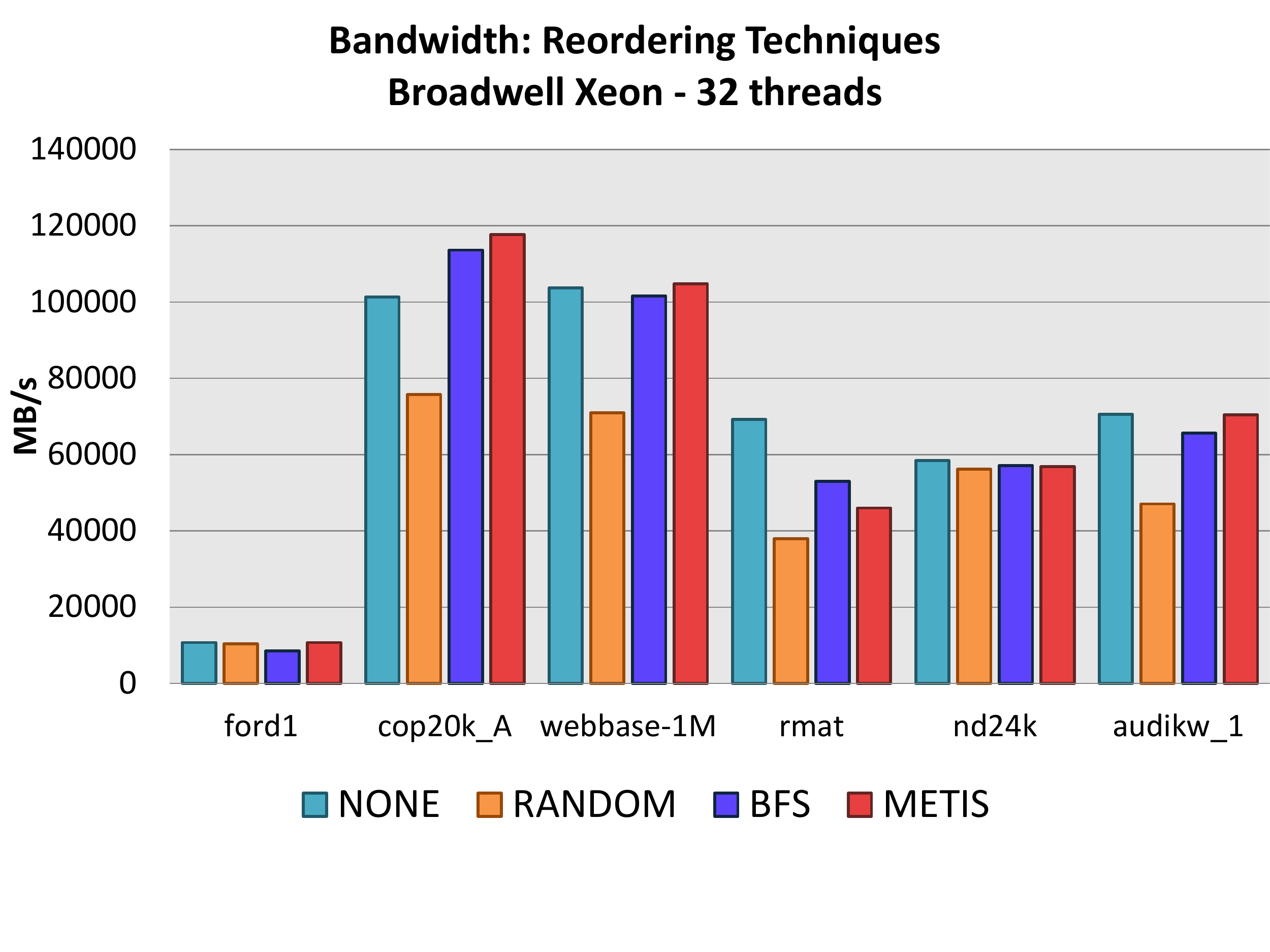}
\caption{SpMV bandwidth (MB/s) on a dual socket Broadwell Xeon system for the different reordering techniques across the matrices when using 32 threads. Higher bars represent better performance. NONE refers to the original, unordered matrix.}
\label{fig:16_Reordering_CPUBW}
\end{figure}

It is worth noting the difference in impact of the matrix reordering techniques on the Emu architecture when compared to a traditional cache-memory based architecture.
Figure \ref{fig:16_Reordering_CPUBW} shows the achieved bandwidth for an identical SpMV implementation on a dual-socket Broadwell Xeon system with 45MB of last-level cache.
We observe that BFS and METIS only achieve up to 12\% and 16\% higher bandwidth over the original matrix, respectively.
Furthermore, a random reordering never outperforms the original matrix, and in general, is considerably worse than the other reordering techniques.
This behavior is what we would expect, as the random reordering has poor spatial locality.
The difference in time to access the L1 cache and main memory can be on the order of 100 -- 200x, which is much more severe than the relative cost of a migration on the Emu architecture.
Such a penalty cannot easily be amortized by the load balancing benefits provided by a random reordering.

\section{Related Work}
\label{sec:relatedWork}
The Emu architecture is described in detail by Dysart \etal~\cite{IA3EMU}, which also gives initial performance results obtained from a simulator of the architecture.
Also using a simulator of an Emu system, Minutoli \etal~\cite{RadixEmu} present an implementation of radix sort and benchmark its performance on up to 128 nodelets.
Hein \etal~\cite{InitialEmu} performed an evaluation of several micro-benchmarks, including CSR-based SpMV, on actual Emu Chick system hardware.
While our work is similar to Hein \etal, there are significant differences. In that work, \textbf{x} was replicated on each nodelet and the entirety of \textbf{b} was placed on a single nodelet.
Furthermore, their evaluation only considered synthetically generated Laplacian matrices.
In our work, \textbf{A}, \textbf{x} and \textbf{b} are all distributed in some fashion across the nodelets, with no portion of the vectors being replicated.
Additionally, we run experiments on real-world sparse matrices that are drawn from a variety of different domains.

\section{Conclusions}
\label{sec:concl}
As migratory thread architectures, such as Emu, mature and evolve, optimizing sparse codes to capitalize on these new architecture's unique strengths will be increasingly important. 
We evaluated several traditional sparse optimizations on the Emu architecture, including vector data layout, work distribution, and matrix reordering.
Our findings can be summarized as follows:
\begin{itemize}
\item While designing data structures and algorithms to minimize migrations is generally a good strategy, we found that work distribution and load balancing is of similar importance for achieving high performance.
\item Unlike traditional systems, it is very difficult to explicitly enforce hardware load balancing for the Emu architecture due to thread migration. Specifically, the placement of the data to be accessed and the patterns of these accesses entirely dictates the work performed by a given hardware resource, irrespective of how much work is initially delegated to each processing element.
\item The impact of employing known matrix reordering techniques is more significant on the Emu architecture than a traditional cache-memory based system. We found that the METIS and BFS matrix reordering techniques can increase performance by as much as 70\% on Emu while we observed a maximum gain of 16\% on a traditional architecture.
Furthermore, a completely random reordering of the rows and columns can exhibit better performance on Emu than not reordering at all, which contradicts what we observe on a traditional system. 
\end{itemize}

For future work, we would like to re-evaluate our performance study on the newly upgraded Emu hardware and toolchain (version 18.08.1),
which includes a faster Gossamer Core clock rate and hot-spot mitigation improvements.
We are also interested in more thoroughly investigating multi-node performance, specifically on hardware, as thus far we have only been able to do so via simulation. 
We would also like to evaluate other sparse matrix formats, including new formats targeted specifically at the Emu architecture.
Furthermore, we are interested in investigating prior work by Valiant~\cite{Valiant} on randomized data distributions and how it can apply to data layout schemes on Emu.  

\bibliographystyle{IEEEtran}
\bibliography{ia32018}

\end{document}